\documentclass{aa}
\usepackage{graphicx}
\usepackage{graphics}
\usepackage{times}
\usepackage{natbib}
\begin{document}
   \title{Foreground removal from \emph{WMAP 5 yr} temperature maps using an MLP neural network}
\titlerunning{Foreground removal by a neural network}

   \author{H. U. N{\o}rgaard-Nielsen \inst{1}}
   \offprints{H. U. N{\o}rgaard-Nielsen}

   \institute{\inst{1}National Space Institute (DTU Space),\\
    Technical University of Denmark, \\
    Juliane Mariesvej 30, DK-2100 Copenhagen, Denmark\\
              \email{hunn@space.dtu.dk}}

   \date{Received ; accepted }

 \abstract
   {}
   {One of the main obstacles for extracting the cosmic microwave background (CMB) signal from observations in the mm/sub-mm range is the foreground contamination by emission from Galactic component: mainly synchrotron, free-free, and thermal dust emission. The statistical nature of the intrinsic CMB signal makes it essential to minimize the systematic errors in the CMB temperature determinations.}
   {The feasibility of using simple neural networks to extract the CMB signal from detailed simulated data has already been demonstrated. Here, simple neural networks are applied to the WMAP 5yr temperature data without using any auxiliary data.}
   {A simple \emph{multilayer perceptron} neural network with two hidden layers  provides temperature estimates over more than 75 per cent of the sky with random errors significantly below those previously extracted from these data. Also, the systematic errors, i.e.\ errors correlated with the Galactic foregrounds, are very small.}
  {With these results the neural network method is well prepared for dealing with the high - quality CMB data from the ESA Planck Surveyor satellite.}


   \maketitle
%

\section{Introduction}

   The cosmic microwave background (CMB) was discovered by Penzias and Wilson (1965). Unique information about the earliest phases of the evolution of the Universe can be derived from CMB temperature and polarization maps. Since its discovery, tremendous effort has been made to improve the CMB maps. Significant improvement has been made with the ongoing NASA \emph{Wilkinson Microwave Anisotropy Probe} (WMAP, Bennet et al. 2003a), and with the Planck mission, launched in May 2009, it is expected that the sensitivity and angular resolution of the CMB maps will be improved by more than an order of magnitude. Unfortunately, the cosmological CMB signal is always mixed with emission from the Milky Way (synchrotron, free-free, and thermal dust emission). To extract the background cosmological information, it is essential to remove the galactic foregrounds without introducing systematic errors.

   Several algorithms have been developed to solve this key issue in CMB research. A comprehensive review is given by Delabrouille and Cardoso (2007). Of course, it is most desirable that the method for removing   the galactic foregrounds produces both a power spectrum and a CMB map with insignificant systematic errors. For the Planck mission, it is an important requirement, since one of the main scientific goals is to search for non-Gaussian features in the CMB maps.

   A lot of signals of individual sky pixels are averaged in order to derive the power spectrum, therefore, the crucial issue is not so much to minimize the random errors per sky pixel, but to minimize the systematic errors in the CMB map as a whole.

   From the FIRAS instrument onboard the COBE satellite, it is known that the CMB spectrum follows a black body spectrum very closely (Mather et al. 1999). Fortunately, all known non-cosmological signals have very different spectral behaviour from a black body. It is thus possible to disentangle the different components of the microwave signals. The obtained accuracy will, of course, depend on the observational errors and frequency coverage of the data available.

   The ESA Planck mission was successfully launched in May 2009, and all systems have been working according to expectations ever since. An important part of the preparation of the mission has been evaluation of the available galactic foreground removal algorithms, based on detailed simulations, called the \emph{Planck Sky Model} (PSM). This work was done by Planck Working Group 2, coordinated by J. Delabrouille and G. de Zotti. Comparisons of the 8 investigated methods can be found in Leach et al.\,(2008).

   N{\o}rgaard-Nielsen and J{\o}rgensen (2008, hereafter NNJ) have shown that with observational errors as expected from the Planck satellite, reasonable assumptions about the spectral behaviour of the galactic foregrounds, it is possible to use simple neural networks to extract the  CMB temperature signal with negligible systematic errors.

   In the analysis of the same PSM data as used by Leach et al.\,(2008), N{\o}rgaard-Nielsen and Hebert (2009, hereafter NNH) have shown that neural networks can also significantly improve the removal of systematic errors in the CMB temperature determination for imaging data.

   An analysis of the WMAP 5yr data is presented here to show the improvement produced by neural networks, also for real observed data. It is basically the same method as in NNJ and NNH, so the neural network references can be found there.

   \section{The CMB data}
   The frequency maps obtained during the first 5 years of the WMAP mission (K, Ka, Q, V, W, centred at 23GHz, 33GHz, 41GHz, 61GHz, 94GHz, respectively) were taken from the official WMAP website:  \emph{lambda.gsfc.nasa.gov/product- /map/current/m\_products.cfm}.

   The PSM maps were taken from the Planck Working Group 2 Challenge-2 ftp area:
   \emph{ftp://planck-wg2.planck.fr/Challenge-2/PSM-maps\_v0}. PSM exposures maps (expected hits per sky pixel) are also provided. In order to derive noise maps for each frequency the algorithm given at the WMAP website has been used, assuming that the noise is Gaussianly distributed. For each of the WMAP frequencies and each of the components (CMB, synchrotron, free-free, thermal and spinning dust) PSM provides maps without observational errors and no corrections for the angular resolution of the different radiometers. To correct for the detector angular resolutions, these maps were convolved with the beam transfer function (simple average per frequency) given on the WMAP website, using the HEALPix synfast routine.

   \section{Modelling the combined foreground spectrum}
   As in NNJ and NNH, the only parameter derived from the observed WMAP 5yr maps is the CMB temperature for each sky pixel.
   In this investigation, only these observed maps have been exploited in order to establish the spectral behaviour of the combined foreground, i.e. no auxiliary data has been incorporated in the analysis.

   To derive the combined foreground maps, the Delabrouille et al. (2008) WMAP 5yr Wiener-filtered map was subtracted from the 5 WMAP frequency
   maps, after the angular resolution was degraded to fit the average beam transfer functions.

   The combined foreground spectra were analysed in terms of the spectral slopes: Ka/K, Q/Ka, V/Q, and W/V, where Ka/K is defined as
   \begin{equation}
   \rm
   Ka/K = log(flux(Ka)/flux(K)) / log(\nu(Ka)/\nu(K)).
   \end{equation}
   and the other slopes are defined in a similar way.

   To ensure that the slopes are determined with reasonable accuracy, only sky pixels close to the Galactic plane are used in the analysis.

\begin{figure}[h]
\centering
\includegraphics[width=3.0 in]{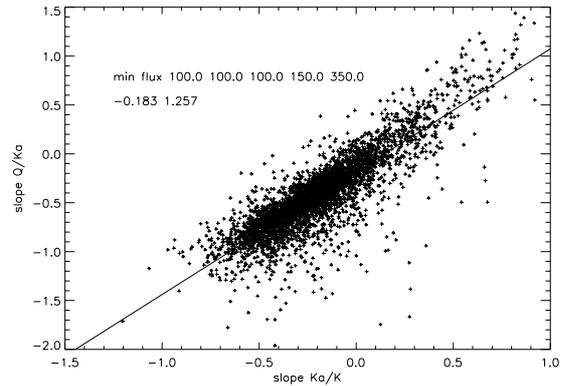}
\caption{The relation between the slopes Ka/K and Q/Ka from the WMAP 5yr maps for sky pixels with minimum fluxes= 100, 100, 100, 150, 350 in the 5 bands. Units: $\rm 10^{-20} erg/cm^{2}/s/Hz/sr$. The linear fit
Q/Ka = -0.183 + 1.257 * Ka/K is shown}
\label{wmap_01_12}
\end{figure}

\begin{figure}[h]
\centering
\includegraphics[width=3.0 in]{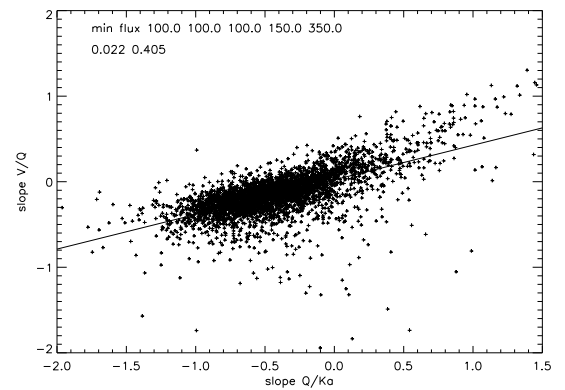}
\caption{The relation between the slopes Q/Ka and V/Q from the WMAP 5yr maps for sky pixels with minimum fluxes= 100, 100, 100, 150, 350 in the 5 bands. Units: $\rm 10^{-20} erg/cm^{2}/s/Hz/sr$. The linear fit
V/Q = 0.022 + 0.405 * Q/Ka is shown}
\label{wmap_12_23}
\end{figure}

\begin{figure}[h]
\centering
\includegraphics[width=3.0 in]{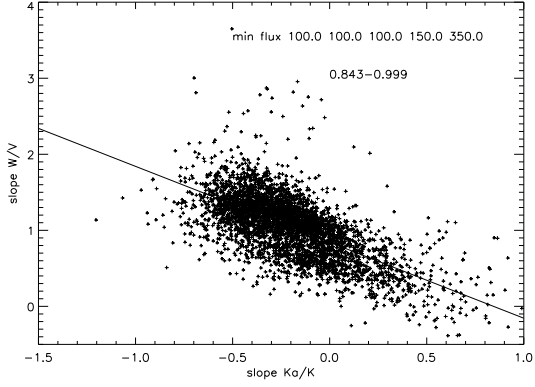}
\caption{The relation between the slopes Ka/K and W/V from the WMAP 5yr maps for sky pixels with minimum fluxes= 100, 100, 100, 150, 350 in the 5 bands. Units: $\rm 10^{-20} erg/cm^{2}/s/Hz/sr$. The linear fit
$W/V = 0.843 - 0.999 * Ka/K$ is shown}
\label{wmap_01_34}
\end{figure}

From Figs. \ref{wmap_01_12}, \ref{wmap_12_23}, and \ref{wmap_01_34} it can be seen that a simple model, which fits the combined foreground spectrum of the WMAP 5yr maps reasonably well, can be derived from linear relations between Ka/K and Q/Ka, Q/Ka and V/Q, and Ka/K and W/V. Therefore, the combined spectrum of the foregrounds used in this investigation has 2 parameters: the flux in the  band and the spectral slope Ka/K
(referred to as NN-foreground).

\section{Brief description of  the neural network concept}

\begin{figure}[h]
\centering
\includegraphics[width=3.0 in]{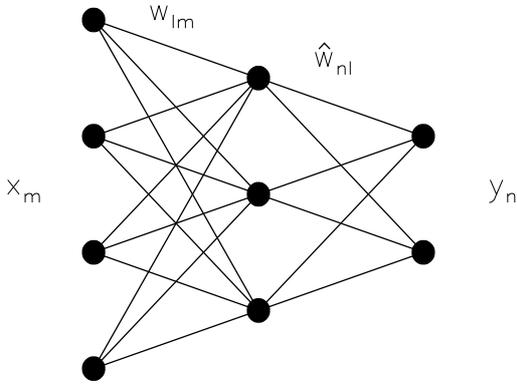}
\caption{A schematic of a multi layer perceptron network with
one hidden layer}
\label{fignnet}
\end{figure}

    Neural networks are analogue computational systems whose structure is inspired by studies of the human brain. An excellent introduction to the many different types of neural networks can be found in Bishop\,\cite{bish95}. In the current paper, as for the previous papers NNJ and NNH, one of the simplest and also most popular networks, the multilayer perceptron (MLP), has been applied. Only a brief description of the neural network method will be given here.

    An MLP consists of a network of units (called processing elements, neurons, or nodes), as illustrated in Fig.~\ref{fignnet}. Each unit is shown as a circle and the lines connecting them are known as weights or links. The network can be understood as an analytical mapping between a set of input variables $x_{m}~(m = 1,...,M)$  and a set of output variables $y_{n}~(n=1,...,N)$. The input variables are applied to the M input units on the left of the figure: M = 4 and N = 2 in the shown example. These variables are multiplied by a matrix of parameters $w_{lm}~(l = 1,...,L);~
    m=1,...,M)$ corresponding to the first layer of links. Here L is the number of units in the middle (hidden) layer: L = 3 in the shown example. This results in a vector of inputs to the units in the hidden layer. Each component of this vector is then transformed by a non-linear function F, giving
    \begin{equation}
    z_{l}~=~F \left( \sum_{m=1}^{M}~w_{lm}x_{m}~+\Theta_{l} \right) ~~(l=1,...,L). \label{eq1}
    \end{equation}
    where $\Theta_{l}$ is an offset or threshold. The Neural Network Toolbox in the MATLAB software environment ($www.mathworks.com$) is used with the $tansig$ function as the non-linear function:
    \begin{equation}
    \mathrm{tansig}(x) ~= ~ \frac{2}{1~+~\mathrm{exp}(-2~x)}~-1. \label{eq2}
    \end{equation}
    It is seen that \emph{tansig} has an S-shape, with values falling within the interval $[-1:1]$.
    From the hidden layer to the output units a linear transformation with weights $\widehat {w}_{nl}~(n=1,...,N;l=1,...,L)$ and offsets $\widehat{\Theta}_{n}$ are applied
    \begin{equation}
    y_{n}~=~\sum_{l=1}^{L}\widehat{w}_{nl}z_{l}~+~\widehat{\Theta}_{n}
		\quad\quad (n=1,...,N). \label{3}
    \end{equation}
    Combining Eqs.\,1 and 2 shows that the entire network transforms the inputs $x_{m}$ to the outputs $y_{n}$ by the following analytical function
    \begin{equation}
    y_{n}(x_{1},...,x_{M})~=~ \sum_{l=1}^{L}\widehat{w}_{nl}~F \left( \sum_{m=1}^{M}w_{lm}x_{m}~+~\Theta_{l}\right)~+~\widehat{\Theta}_{n}. \label{eq4}
    \end{equation}
    Clearly, such an MLP can be easily generalized to more than one hidden layer.

    Given a set of P example input and output vector pairs $\{x_{m}^{p}~y_{n}^{p}\}~ p=1,...,P$ for a specific mapping, a technique known as error back propagation, can derive estimates of the parameters $w_{lm},~\Theta_{m}$ and   $\widehat{w}_{nl},~\widehat{\Theta}_{n}$, so that the network function (\ref{eq4}) will approximate the required mapping.
    The training algorithm minimizes the error function
    \begin{equation}
    E_{NET} ~=~\sum_{p=1}^{P}\sum_{n=1}^{N}[y_{n}(x^{p}) ~- ~y_{n}^{p}]^{2}. \label{eq5}
    \end{equation}

    A neural network is set up to handle a given data set. Traditionally, this is split into 2 data sets: one used directly to train the network and a validation data set used in the iteration scheme, not directly in the training, but in the evaluation of the improvement of the network.  Furthermore, a test data set that is only used at the end of the training to get an independent estimate of the accuracy of the derived network.

\section{The applied neural network}
A basic assumption for this method is that the noise is white (i.e. no 1/f noise). If this is not the case, it is necessary to correct the maps for non - white features, before the data is run  through the network. It is beyond the scope of this paper to discuss this issue in relation to the WMAP 5yr data. With the assumption of white noise, the noise of the individual sky pixels is independent, and it is possible to treat each pixel separately.

The neural networks applied here have 5 input channels, one for each of the WMAP 5 frequencies and one output channel, the CMB temperature. Together the 5 input values are referred to as a spectrum. The setup of the neural network follows this scheme:
\begin{enumerate}
\item For each simulated spectrum in the data set used to set up the neural network, draw 2 uniformly distributed numbers, one within the range covered by the K flux in the WMAP 5yr map, and one covering the range of the spectral slope Ka/K as seen in Fig. 1.

\item Calculate the resulting spectrum for the 5 WMAP bands from the linear slope relations in Figs. \ref{wmap_01_12}, \ref{wmap_12_23}, and \ref{wmap_01_34}.

\item Draw a uniformly distributed number within the range of temperatures found previously in WMAP 5yr CMB maps, and add the black body spectrum to the spectrum calculated in step 2.

\item For each frequency, add a random Gaussian noise corresponding to the noise in the observed WMAP 5 yr maps.

\item Repeat 1--4 until the desired number of spectra ($N_{NNET}$) has been obtained. This data set is split into a set used directly to train the network and a set used for validation of the iteration scheme.

\item Train the neural network to find the transformation between the input spectra and the true CMB temperatures (known for each spectrum of the training data set).

\item Obtain an independent test sample of spectra by repeating 1--4 $N_{TEST}$ times

\item Run the $N_{TEST}$ spectra through the network to get an independent estimate of the reliability of the network

\item If the derived network is working satisfactorily on the test data set, run the WMAP 5yr data through the network

\end{enumerate}

 An MLP with 2 hidden layers (5 and 3 processing elements, respectively, referred to as the NN network) was used for the data sets considered here.
The experience is that about 10000 spectra is enough for the data set used to train the network.

\begin{figure}[h]
\centering
\includegraphics[angle=90, width=3.0 in]{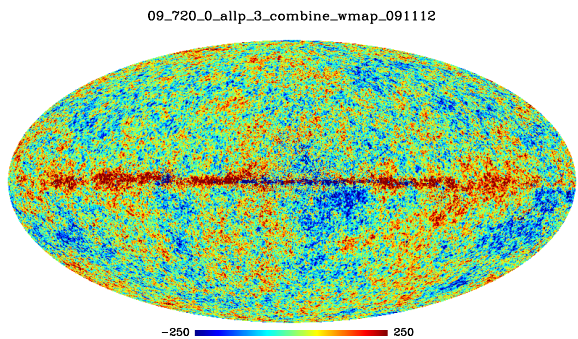}
\caption{The CMB map extracted from the WMAP 5 yr maps with the [5,3] neural network (units: $\mu$K)}
\label{hunn_wmap}
\end{figure}

\section{Results of the neural networks}

Figure~\ref{hunn_wmap} gives the CMB temperature maps extracted from the WMAP 5 yr maps by NN.
For an estimate of the systematic errors in this map, a set of 30 maps were constructed from the NN-foreground model including Gaussian noise derived from the WMAP hitmaps.

For one of these sets of simulated maps, Fig.~\ref{09_720_0_model_128} shows the residual $(T(in) - T(out))$ map as output from NN.
The neural network has a problem covering the full dynamic range of the fluxes close to the Galactic plane. The inner part of the Milky Way has been masked out (WMAP KQ75 mask = 0). The errors are close to being Gaussianly distributed (skewness = 0.00, kurtosis = 0.09), with only very small systematic errors.

\section{Previous WMAP 5yr CMB maps}

The data released by the WMAP team was extensively analysed. Delabrouille et al. (2009, hereafter DCLBFG) discuss the different methods applied to extract the CMB signal in detail.

The WMAP team has extracted `internal linear combination' (ILC) maps, simply obtained by a summation of the 5 frequency maps (after reduction of the resolution of the maps to 1 deg). The weights for each map are determined by  minimizing the total variance of the output map. The sky is divided into 12 areas, in which the weights are estimated independently. The KQ75 mask, used in this investigation, is contained within Area 0. A basic problem with the ILC method is  that it does not take the known variations of the spectral shapes and the relative contributions of the different foreground components into account. Due to the statistical properties of the ILC map, the WMAP team does not recommend using it for cosmological investigations.

Only Tegmark et al.\,(2003, hereafter TOH) and DCLBFG have produced CMB maps with an angular resolution close to resolution of the best, W, channel. TOH developed a variant of the ILC method in the spherical harmonic domain, in which the weights are allowed to vary as a function of the multipole l. The weights are computed in 9 independent areas on the sky. The method was developed by means of the WMAP 1 yr maps and the WMAP 5yr maps have been analysed with the same method and made available at
\emph{http://space.mit.edu/home/tegmark/cleaned5yr.map.fits}.

The DCLBFG team developed a variant of the ILC method exploiting the properties of needlet functions (Narcowich et al. 2006). These functions are both localized in the spherical harmonic domain and in the spatial domain. They incorporated the IRAS IRIS 100 $\mu$m map in the analysis ( Miville-Desch\^{e}nes and Lagache, 2005). All WMAP maps are deconvolved to the resolution of the W channel, which, of course, has increased the noise in the maps with poorer angular resolution.

DCLBFG used maps produced with the PSM simulation package to check the level of systematic errors in their CMB map. In their Fig. 4 they give the residual map: some structures are seen but are difficult to quantify since the colour scale of the map covers $\pm$ 600 $\mu$K.

The DCLBFG ILC map has been taken from \emph{http://- www.apc.uni-paris7.fr/APC.CS/Recherche/Adamis/cmb.wmap-en.php}

\begin{figure}[h]
\centering
\includegraphics[angle=90, width=3.0 in]{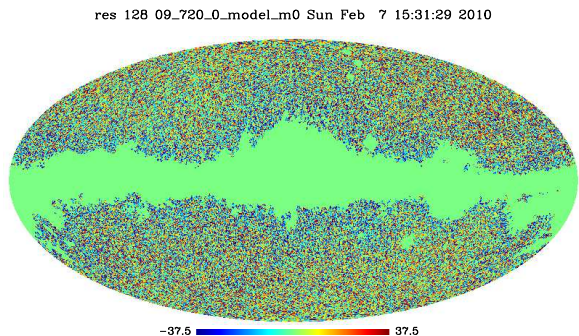}
\caption{The residual map $(T(in) - T(out))$ of the neural network fitting of a typical map derived from the NN-foreground + CMB model (units: $\mu$K).
The green area covering the inner part of the Milky Way is the WMAP KQ75 mask = 0. The map contains no systematic errors outside this area and the resolution of the map is HEALPix nside = 128}
\label{09_720_0_model_128}
\end{figure}

\begin{figure}[h]
\centering
\includegraphics[angle=90, width=3.0 in]{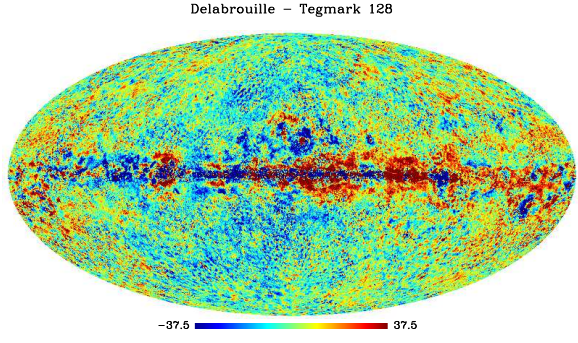}
\caption{The difference between the CMB maps by Delabrouille et al. (2009) and Tegmark al. (2003) (unit: $\mu$K).
The HEALPix nside  and the colour scale are the same as  Fig.\ref{09_720_0_model_128}}
\label{res_dela_tegm_128}
\end{figure}

The difference between the DCLBFG and the TOH CMB maps is given in Fig.\ref{res_dela_tegm_128} . Due to an apparent  small difference in temperature scale, the TOH map has been multiplied by a factor of 1.04. The scale and the HEALPix nside is the same as in Fig.\ref{09_720_0_model_128} (i.e.\,$\pm$ 37.5 $\mu$K, 128). The residuals in Fig.\ref{res_dela_tegm_128} deviate from a Gaussian distribution (skewness = 0.00, kurtosis = 0.26), and the systematic errors seem to reach a level of $\pm$ 20 $\mu$K inside the KQ75 mask.

Both the NN map in Fig.\ref{hunn_wmap} and the DCLBFG and TOH maps have obvious problems estimating the CMB temperatures close to the Galactic plane. Therefore, only the sky area covered by the WMAP KQ75 mask will be used in deriving the power spectra of the maps, see below.

As emphasized above, the WMAP team has not produced a CMB map where it is possible to derive their optimal power spectrum directly. The method for deriving the optimal power spectrum is described in detail in Hinshaw et al\,(2003). Briefly, it is based on the maps of each individual V and W detector, cleaned by foreground templates, and using the cross power spectra for each combination of spectra.

\begin{figure}[h]
\centering
\includegraphics[width=3.0 in]{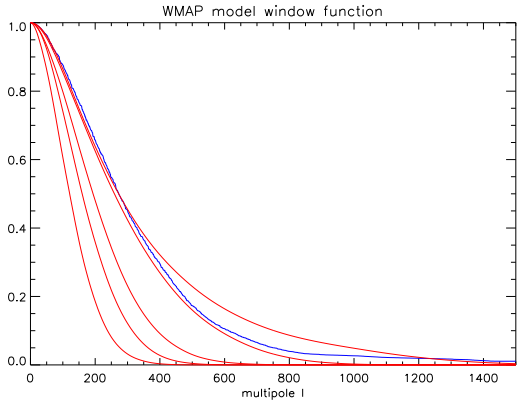}
\caption{The window transfer function (blue curve) obtained from the average of the power spectra of the 30 sets of simulated CMB maps. The averages of the individual window functions, given by the WMAP team, are shown as red curves. It is seen that the angular resolution of the NN maps are somewhere between the resolution of the V and W channels}
\label{model_window}
\end{figure}

As seen above, NN uses maps with different angular resolutions. To estimate the effective NN window function, 30 sets of simulated NN-foreground  + CMB maps were run through NN, and the window function was derived from the average power spectrum of these 30 CMB maps (see Fig.\ref{model_window}). It is seen that the angular resolution of the NN CMB maps is somewhere between the resolution of the V and W channels.

\begin{figure}[h]
\centering
\includegraphics[width=3.0 in]{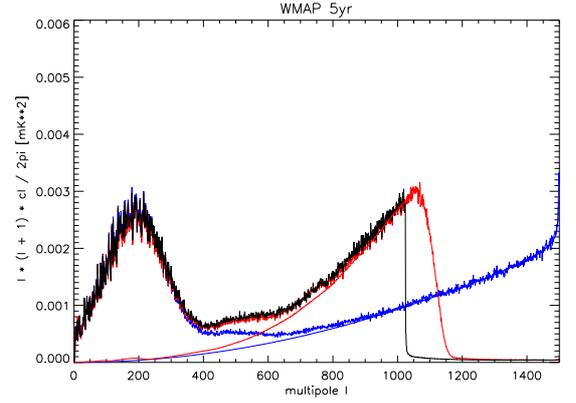}
\caption{The power spectra of the NN CMB maps (blue curve), the Delabrouille et al.\,(2009) map (red curve), and the Tegmark et al.\,(2003) map (black curve).
The power spectra of the noise in the NN map and the Delabrouille et al.\,(2009) map are given. }
\label{model_pow_1}
\end{figure}

\begin{figure}[h]
\centering
\includegraphics[width=3.0 in]{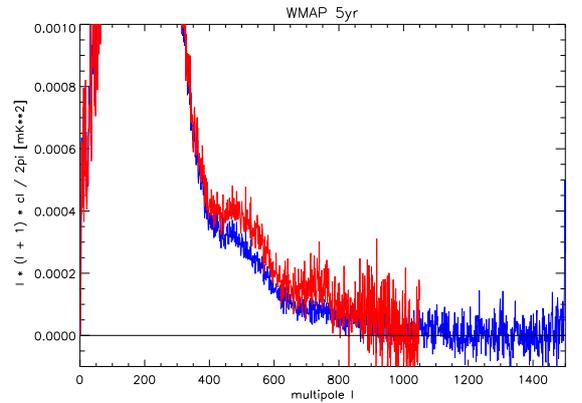}
\caption{The NN CMB map (blue curve) and the Delabrouille et al. (2009, red curve) map with the noise power spectra subtracted. }
\label{model_pow_corr_1}
\end{figure}

\begin{figure}[h]
\centering
\includegraphics[width=3.0 in]{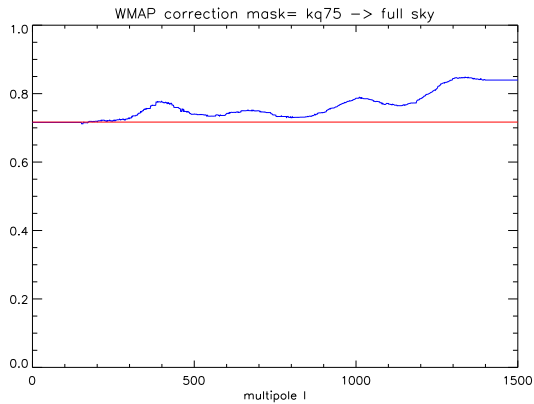}
\caption{The correction to the power spectra of the CMB maps analysed in this paper, to compensate for the sky coverage of the KQ75 mask (geometrically $\sim$71 percent of the sky, seen as the horizontal line). The correction has been derived from 30 realizations of the PSM reference CMB map, as the mean of the ratio between the power spectra obtained with and without applying the KQ75 mask. Furthermore, the correction has been median smoothed with $\rm \Delta l$ = 25}
\label{corr_pow_kq75_full}
\end{figure}

\begin{figure}[h]
\centering
\includegraphics[width=3.0 in]{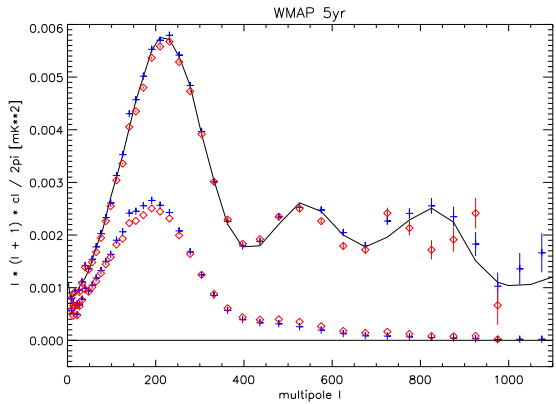}
\caption{The power spectra of the NN (blue crosses) and the Delabrouille et al. (red diamonds) CMB maps, with and without corrections for the beam and pixel window functions and the KQ75 sky coverage.
The black line is the best $\rm \Lambda$CDM model derived by the WMAP team for the 7yr maps(Larson et al. 2010) }
\label{model_pow_window_2}
\end{figure}

\begin{figure}[h]
\centering
\includegraphics[width=3.0 in]{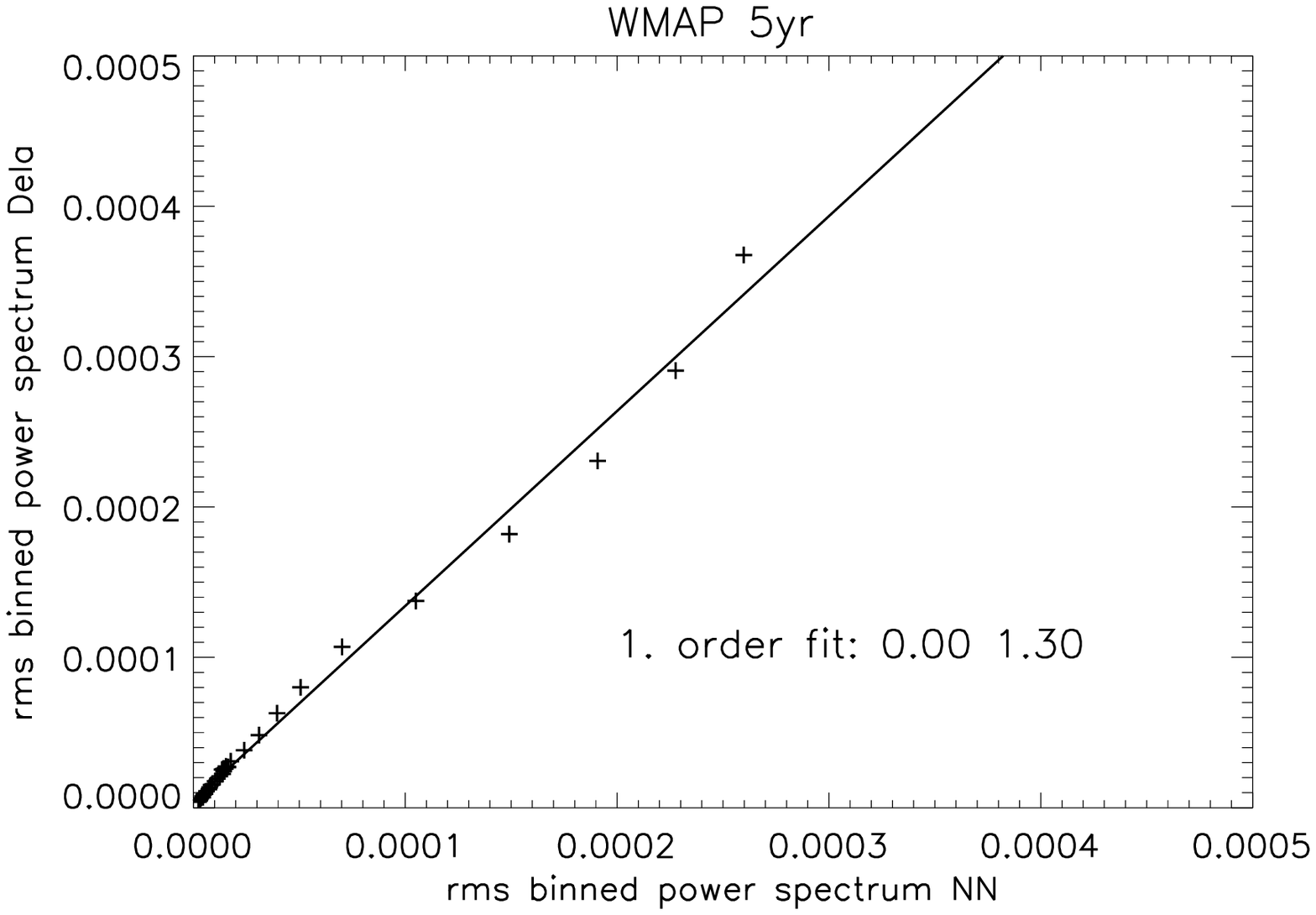}
\caption{The relation between the rms of the binned power spectra shown in Fig. \ref{model_pow_window_2} (X-axis: NN, Y-axis: DCLBFG), indicating that the rms of the Delabrouille et al. binned spectrum is about 30 per cent greater than for the NN spectrum }
\label{rms_corr_pow}
\end{figure}

The power spectra of the WMAP 5yr CMB maps by NN , DCLBFG,  and TOH are shown in Fig. \ref{model_pow_1}.
The noise power spectrum of NN has been established by means of 100 simulated noise maps. The DCLBFG noise power spectrum for the full sky is taken from the website. In DCLBFG Fig. 5, the derived noise power spectra of the sky above and below Galactic latitude 15 degrees are given, the main difference being the noise level around the first Doppler peak. Since the noise power spectrum of the whole sky is quite similar to the $\rm |b| >$ 15 degrees power spectrum, the full sky noise spectrum scaled to the power spectrum in the KQ75 mask around l = 950 has been used in the following. Since TOH are not giving detailed information about the noise spectrum in their CMB map, this map is not analysed further here. In Fig. \ref{model_pow_corr_1} the noise spectra were subtracted from the NN  and DCLBFG power spectra. It is evident that the angular resolution of the DCLMFG map is better than the NN map.

To estimate the errors in the NN and DCLBFG power spectra due to the WMAP observational errors, a series of 30 sets of NN-foreground + CMB maps, including noise at the appropriate level from Fig 9, were run through the NN network . At each l-value, the rms of a  power spectrum was taken as the rms around the mean power spectrum of the simulated maps. The error bars have been derived from these rms estimates, with the definition of multipole ranges given by the WMAP team (Nolta et al. 2009).

The power spectra were corrected for the beam and pixel window functions, and for the sky coverage of the KQ75 mask.
The pixel window function was derived from the reference power spectrum and a realization map with nside = 512.
To correct for the sky coverage of the KQ75 mask, the power spectra of 30 realization of the PSM reference CMB map (the optimal $\Lambda$CDM fit to the WMAP 1yr power spectrum by the WMAP team was obtained both with and without applying the mask. Since these power spectra are only used in a relative way, this reference map is adequate for the purpose. The mean ratio , median - filtered with $\rm \Delta l = 25$, is given in Fig. 11.
For the NN spectrum, the beam window function given in Fig. \ref{model_window} was used, while for the DCLBFG spectrum, the window function supplied by DCLBFG (equal to the function provided by the WMAP team up to l = 1050) was applied.

The corrected NN and DCLFBG power spectra are given in Fig. \ref{model_pow_window_2}. The best - fit $\Lambda$CDM spectrum derived by the WMAP team from the WMAP 7yr maps (Larson et al. 2010) is also shown. It is seen that the NN spectrum follows this theoretical spectrum reasonable well out to l = 1100. The rms of the binned power spectra in Fig. \ref{model_pow_window_2} is plotted against each other in Fig. \ref{rms_corr_pow}, indicating that the accuracy of the DCLBFG power spectrum is about 30 percent poorer than for the NN spectrum.

\section{Conclusions}

The feasibility of a neural network for extracting the CMB signal from mm/sub-mm observations was extended to real observations, namely the WMAP 5 yr maps, without adding any auxiliary data. It was demonstrated that an MLP neural network with 2 hidden layers can extract the CMB map with random errors significantly smaller than those previously obtained for this data set, for about 75 per cent of the sky. The systematic errors---errors correlated with the Galactic foregrounds---are also very small, and the power spectrum was extended up to l = 1100. It fits the optimal $\rm \Lambda$ CDM model obtained by the WMAP team from the 7yr maps well.

To fully qualify the neural network method for the Planck data analysis, a forthcoming paper will demonstrate the feasibility of extracting the CMB polarization signal using the detailed simulations in the Planck Sky Model.

\acknowledgements
We would like to thank Dr.\,C.\,A.\,Oxborrow for significant improvements to the language and typesetting of this paper. We acknowledge the use of the Planck Sky Model developed by the Component Separation Working Group of the Planck Collaboration.

\end{document}